\documentclass[10pt,letterpaper,twocolumn]{article} 

\usepackage{ol2}
\usepackage[draft]{hyperref}
\usepackage{amsmath}

\begin{document}

\twocolumn[ 

\title{Experimental generation of ring-shaped beams with random sources}


\author{Salla Gangi Reddy$^*$, Ashok Kumar$^1$, Shashi Prabhakar, and R. P. Singh}

\address{
Physical Research laboratory, Navarangpura, Ahmedabad, India-380 009. \\
$^*$Corresponding author: sgreddy@prl.res.in \\ 
$^1$Presently at Instituto de Fisica, Universidade de São Paulo, São Paulo, 66318, Brazil.
}

\begin{abstract} 
We have experimentally reproduced ring shaped beams from the scattered Laguerre-Gaussian and Bessel-Gaussian beams. A rotating ground glass plate is used as a scattering medium and a plano convex lens collects the scattered light to generate ring shaped beams at the Fourier plane. The obtained experimental results are supported with the numerical results and are in good agreement with the theoretical results proposed by Wang, Cai and Korotkova (Opt. Exp. \textbf{17,} 22366 (2009)).
\end{abstract}


 ] 

Ring-shaped or dark hollow beams have found applications in guiding cold atoms \cite{cold} and trapping of low refractive index particles \cite{low}. Such beams can be generated through multi-mode light wave guides \cite{hallow}, spiral phase plates \cite{spp}, multi-mode fibers \cite{mmf} and computer generated holograms (CGH) \cite{cgh}. 

Optical vortices or the Laguerre-Gaussian (LG) beams with zero radial index have attracted a great deal of attention due to their applications in optical manipulation, optical communication and quantum information \cite{OAM, torner}. These beams are recognized with a dark core in their intensity profile and a helical wavefront \cite{core,shashi}. We have considered LG beams with zero radial index throughout the paper. Along with the LG beams, Bessel beams owing to their interesting properties of propagation without an apparent spreading due to diffraction have also been a subject of study since more than two decades \cite{bessel}. Usually in a laboratory the Bessel beams are generated using a Gaussian laser beam and termed as Bessel-Gaussian (BG) beams. In this letter, we demonstrate the generation of ring-shaped beams from the scattered LG and BG beams.

The scattered light of a Gaussian laser beam through a rotating ground glass (RGG) plate can be modeled as a Gaussian Schell-model (GSM) beam \cite{mmf}. This GSM beam is partially coherent light which has Gaussian intensity distribution and Gaussian spectral degree of coherence. Recently a lot of applications of partially coherent beams have been suggested in diverse areas \cite{cai}. Wang et al \cite{korotkovaoe} have introduced the partially coherent LG beams of all orders. The  temporal coherence properties of partially coherent beams generated by the scattering of optical vortices through a RGG plate have also been studied \cite{ashok,ashok1}. It has been shown that the decay of coherence becomes sharper with increase in the order of LG beam incident on the RGG plate. A similar type of behavior has been observed theoretically in the Fourier transform of the spatial correlation function of the LGSM and the BGSM beams \cite{olga}. It has been stated that the beams having a rotational symmetry in the Fourier transform of their spatial correlation function and zero value on the beam axis can generate a dark core in the far field intensity distribution. Mie and Korotkova \cite{olga} generated ring shaped beams with an arbitrary beam (including Gaussian beam) by introducing LG correlation function through a phase screen. We have experimentally generated ring shaped beams with LG (BG) beams by introducing Gaussian correlation function through a RGG plate. The obtained experimental results are simulated by using the expression of cross spectral density of the partially coherent beams generated by a Schell model source, and at $z=0$ it is given by \cite{korotkovaoe}
\begin{align}
 W(x_1, y_1, x_2, y_2, 0)= \sqrt{I_1(x_1, y_1,0) I_2(x_2,y_2,0)} \nonumber \\  g(x_1-x_2,y_1-y_2,0)
\label{oe}.
\end{align}

where $I_1(x_1, y_1,0)$ and $I_2(x_2,y_2,0)$ are the intensity distributions at the positions $(x_1, y_1,0)$ and $(x_2, y_2,0)$ respectively; $g(x_1-x_2,y_1-y_2,0)$ is the spectral degree of coherence.
  
\begin{center}
   \begin{figure}[htb]
   \includegraphics[width=8.4cm]{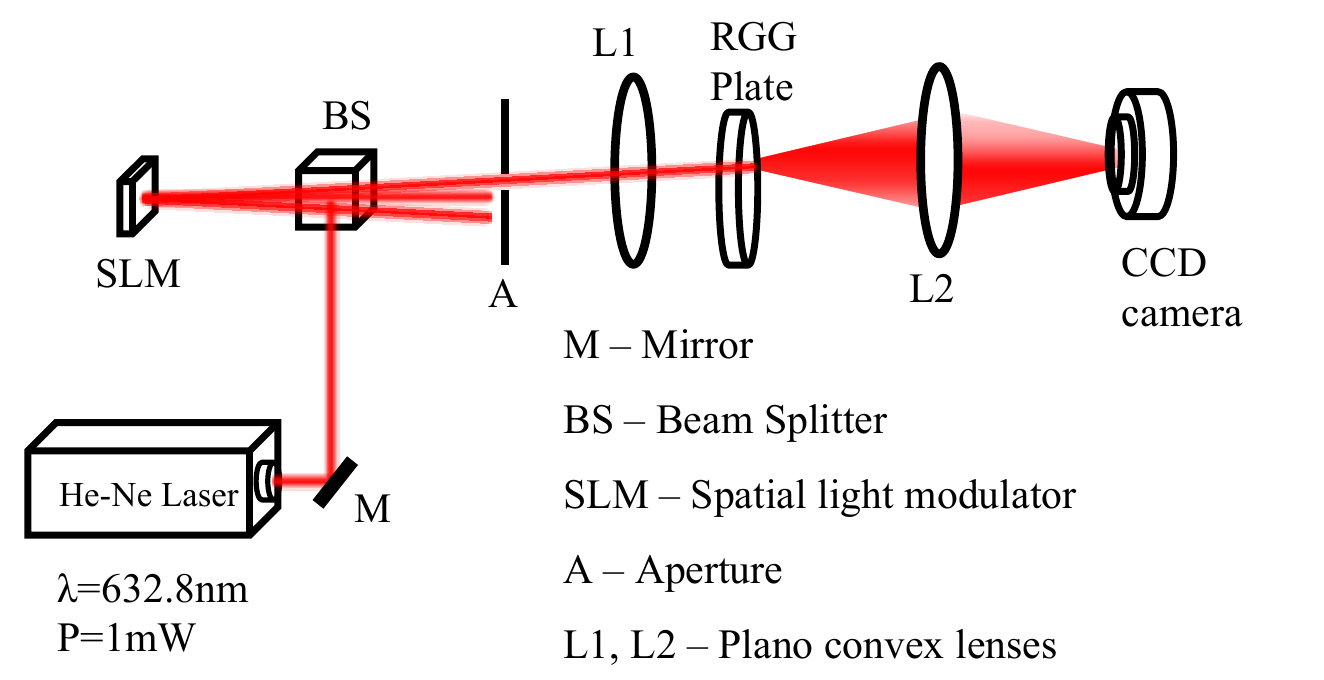}
    \caption{(Colour online) Experimental set up for the generation of ring shaped beams.}
    \label{fig:expt}
   \end{figure}
   \end{center}

 Our experimental set up for the generation of ring-shaped beams is shown in Fig. \ref{fig:expt}. An intensity stabilized He-Ne laser beam of maximum power 1 mW and beam waist 0.3 mm is used to generate LG and BG beams. These beams are produced with the computer generated holography technique using a spatial light modulator (SLM). Different computer generated holograms for generating different LG and BG beams are introduced to the SLM through a computer. The required beam is selected with an aperture $A1$, and passed through the lens (L1) of focal length 25 cm and the RGG plate.The RGG plate is translated along the direction of incident beam to change the width of this beam on the plate. We have done experiment for four positions of the RGG plate by translating in steps of 2.5 cm. The scattered light from the RGG can be approximated as the corresponding Schell-model beam which is focused with a plano convex lens (L2) of focal length 10 cm. The images corresponding to the different input beams are recorded with a CCD camera. The SLM is placed at a distance of 63 cm from the laser and the lens L1 is placed at 56 from the SLM. The lens L2 is kept at a distance of 24 cm from the RGG plate and the CCD camera is placed 37 cm from the lens. Position of the lens L2 is adjusted to get a clear far field intensity distribution for an optimum diameter of the ring shaped beams that could be captured with the CCD camera being used by us.  
 
\begin{figure}[htb]
      \begin{center}
      \includegraphics[width=8.4cm]{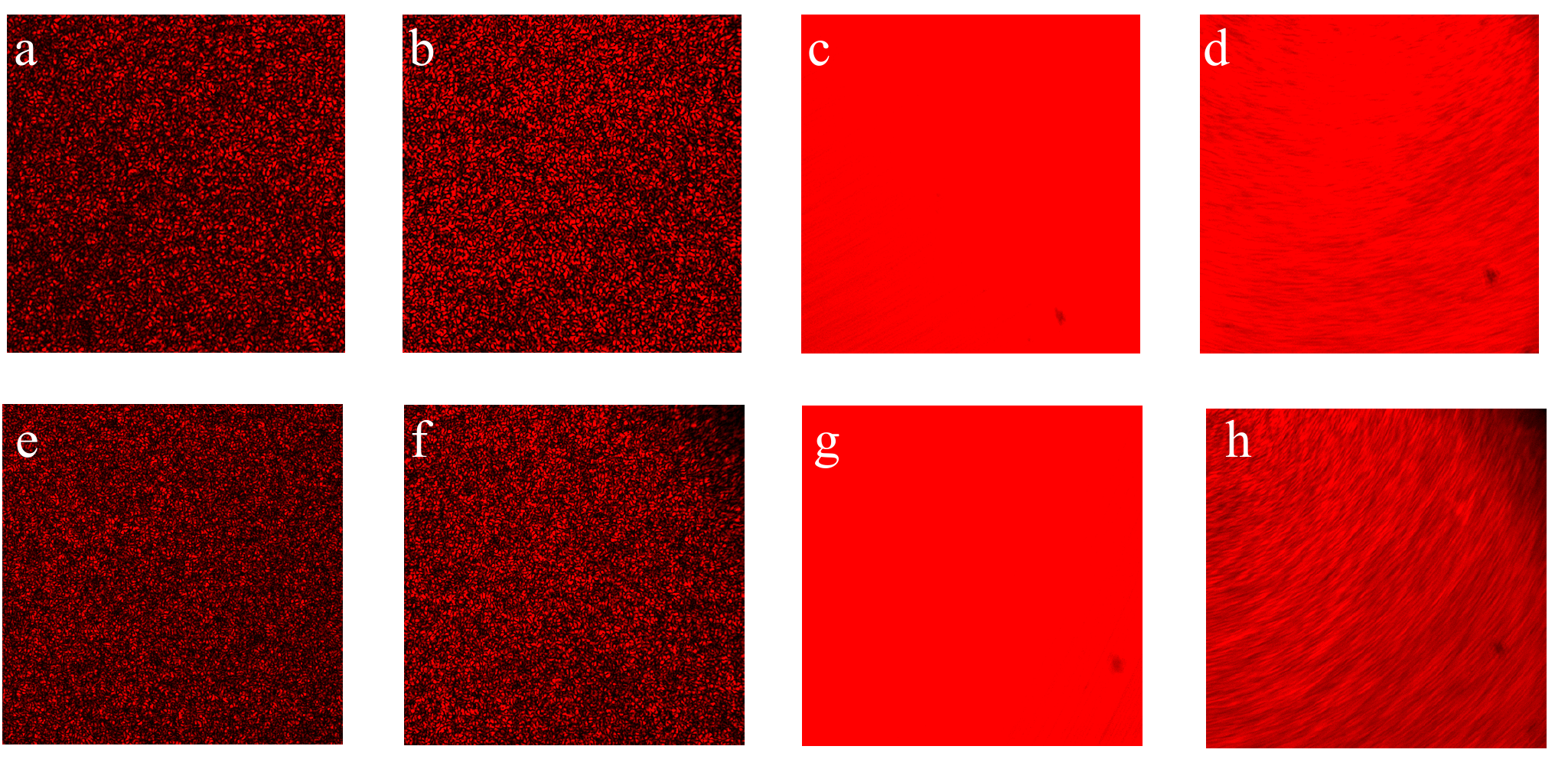}
       \caption{(Colour online) Images showing the intensity distributions of scattered second order LG (a-d) and BG (e-h) beams; (a, e) are recorded after the ground glass and (b, f) after the lens, while (c, g) and (d, h) are recorded at same places when the ground glass is rotating (linear speed 72.1 cm/sec).}
       \label{fig:BG02_DIF} \vspace{0.1cm}
        \end{center}
      \end{figure}

 We start our experiment with the recording of the images of scattered second order LG and BG beams (azimuthal index 2) from the static ground glass plate. The images are captured at the distance of 5 cm from the plate and also at the 18 cm from the focusing lens (L2); both for a incident beam of width 1.1 mm. The same are also recorded with the rotating ground glass plate. These images are shown in Fig. \ref{fig:BG02_DIF}. One can notice that the recorded images do not show any intensity distribution like original LG and BG beams. The random intensity distributions obtained for static ground glass (Fig. 2(a,b,e,f) gets averaged out in case of the RGG plate (Fig. 2(c,d,g,h)).
 
\begin{figure}[htb]
     \begin{center}
     \includegraphics[width=8.4cm]{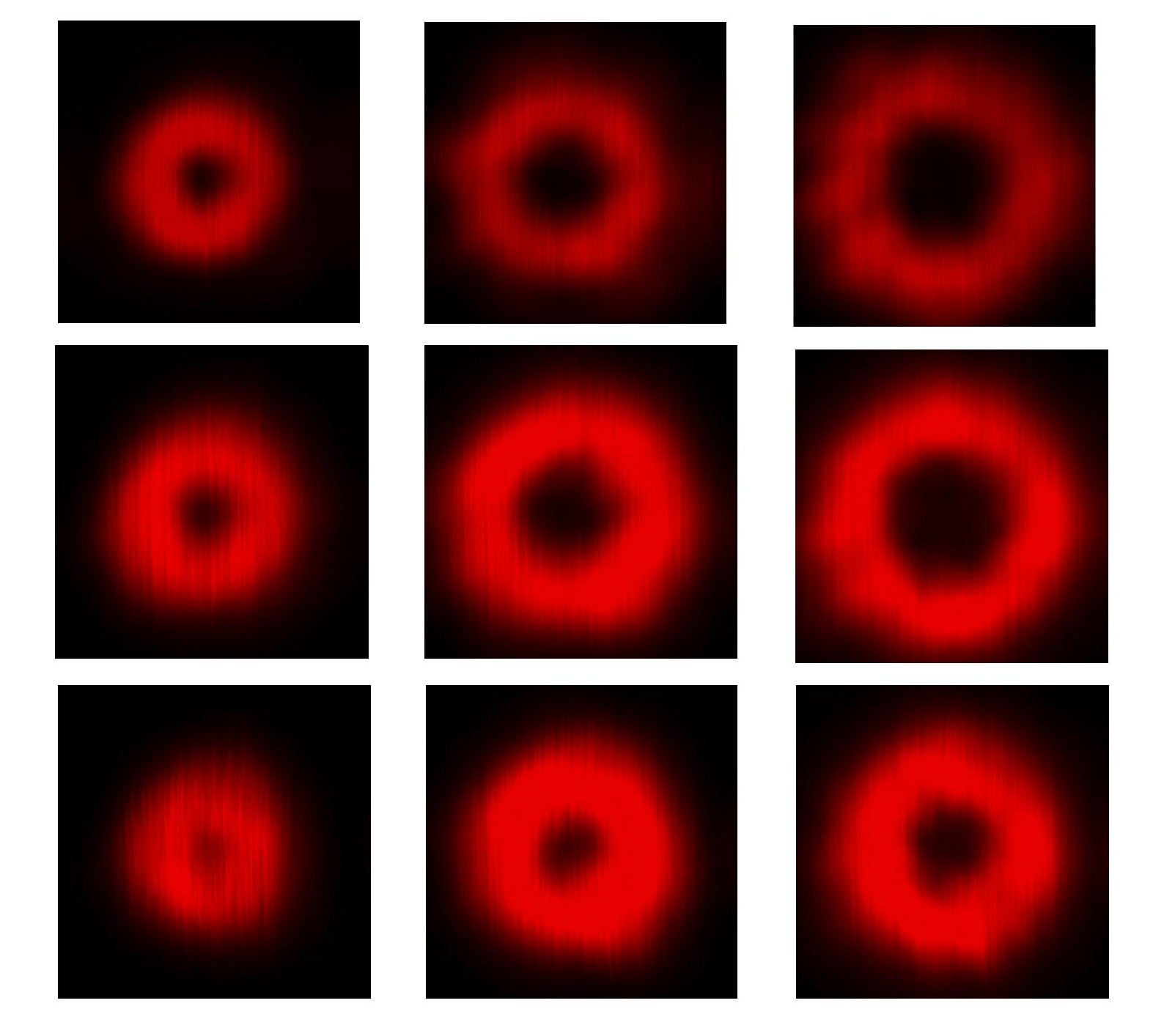}
      \caption{(Colour online) Far field intensity distribution of the scattered LG beams of different azimuthal indices ($\rm m = 2, 4, 6$) through a RGG plate for different widths of the incident beam, 0.496 mm (top), 0.412 mm (middle) and 0.321 mm (bottom).}
      \label{fig:lgsm}
      \end{center}
     \end{figure}

\begin{figure}[htb]
    \begin{center}
    \includegraphics[width=8.4cm]{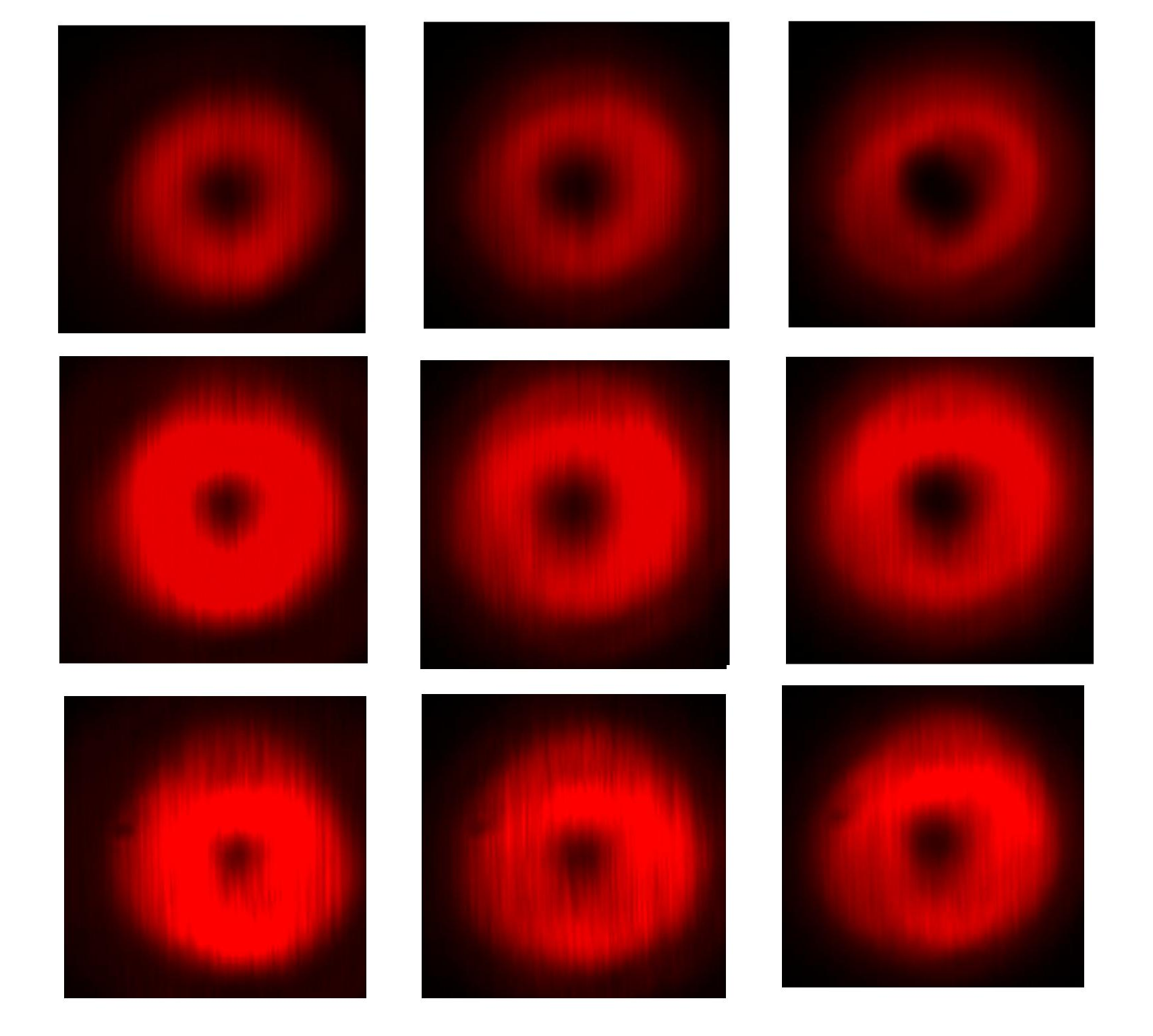}
     \caption{(Colour online) Far field intensity distribution of the scattered BG beams for same conditions as in Fig. \ref{fig:lgsm}}
     \label{fig:bgsm}  \vspace{0.1cm}
     \end{center}
    \end{figure}

\begin{figure}[htb]
    \begin{center}
    \includegraphics[width=8.4cm]{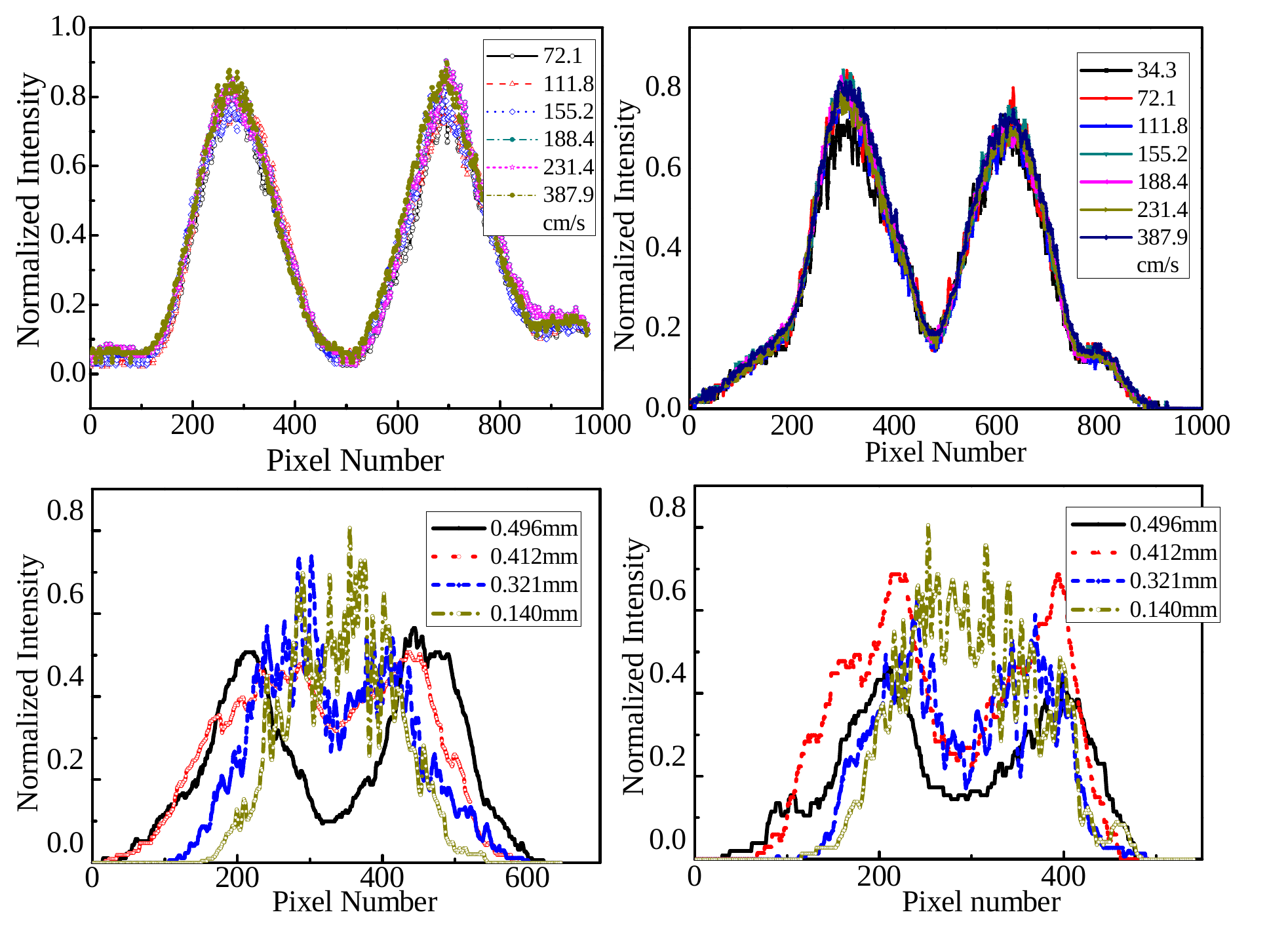}
     \caption{(Colour online) The line profiles (intensity distribution) along the core of ring shaped beams generated from scattered second order LG (left) and BG (right) beams at different speeds of the RGG plate (top) and for different incident beam widths (bottom)}
      \label{fig:line2}
      \end{center}
    \end{figure}

The far field intensity distributions of the scattered LG and BG beams have respectively been shown in Figs. \ref{fig:lgsm} and \ref{fig:bgsm}. It could be seen that the far field intensity distributions form ring shaped beams with dark core for incident beams with nonzero azimuthal index. We have shown the far field intensity distribution of scattered LG and BG beams with azimuthal indices 2, 4 and 6 (radial index is zero for all images) for a speed 34.3 cm/s of the RGG plate. The diameter of the dark core increases with increase in the azimuthal index for both the LG and the BG beams. Since it is well known that the spatial coherence of the scattered light from the RGG plate is proportional to the incident beam width \cite{spatial}, therefore to study the effect of spatial coherence of the scattered light on far-field intensity distributions, we change the width of the incident beam on the RGG plate. We have observed that the dark core disappears at very low coherences (if the beam width is less than 0.140 mm for first order scattered vortex). We have also observed that the diameter of dark core is independent of the speed of the RGG plate i.e. temporal coherence of the scattered light. This has been shown by drawing the line profiles along the dark core of the far field intensity distributions of scattered second order LG and BG beams at different speeds of the RGG plate for incident beam width of 1.1 mm, shown in Fig. \ref{fig:line2}.

\begin{figure}[htb]
 \begin{center}
 \includegraphics[width=7.4cm]{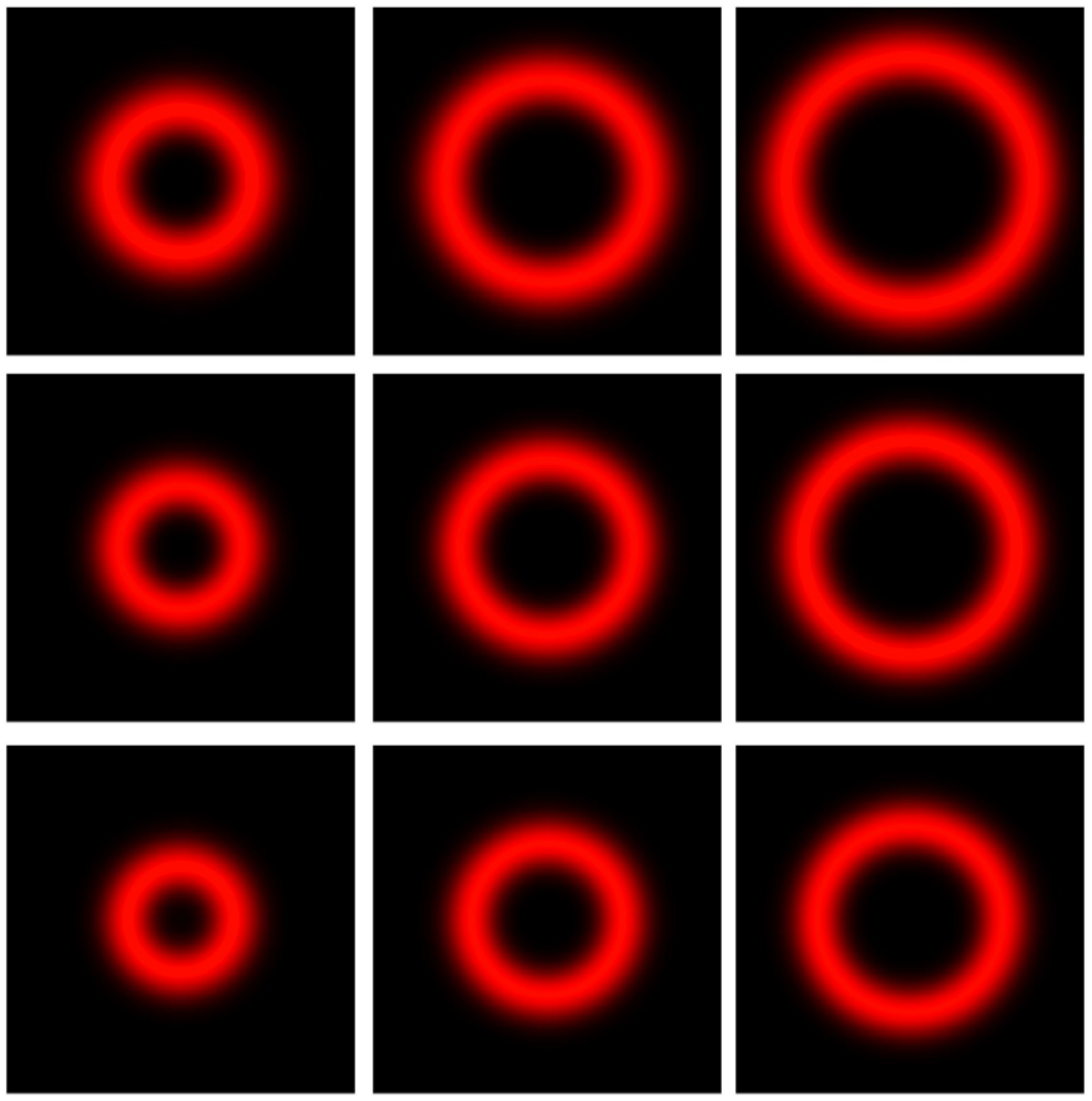}
  \caption{(Colour online) Theoretical results for far field intensity distribution of the scattered LG beams of different azimuthal indices ($\rm m = 2, 4, 6$) through a RGG plate for different $\sigma$ values in \cite{korotkovaoe}, 0.496 \rm{mm} (top), 0.412 \rm{mm} (middle) and 0.321 \rm{mm} (bottom).}
  \label{fig:lgsm1}
  \end{center}
  \end{figure}

 We have also simulated the dependence of the presence of dark core on the width of incident beam or the spatial coherence of the beam; shown in Fig. \ref{fig:lgsm1}. These theoretical results have been obtained by using Eq. (10) of \cite{korotkovaoe}. We have used the following ABCD matrices for the propagation of partially coherent beam through free space of distance $z_1$, lens of focal length $ f $ and free space of distance $z_2$. 
  
\begin{figure}[htb]
      \begin{center}
      \includegraphics[width=8.4cm]{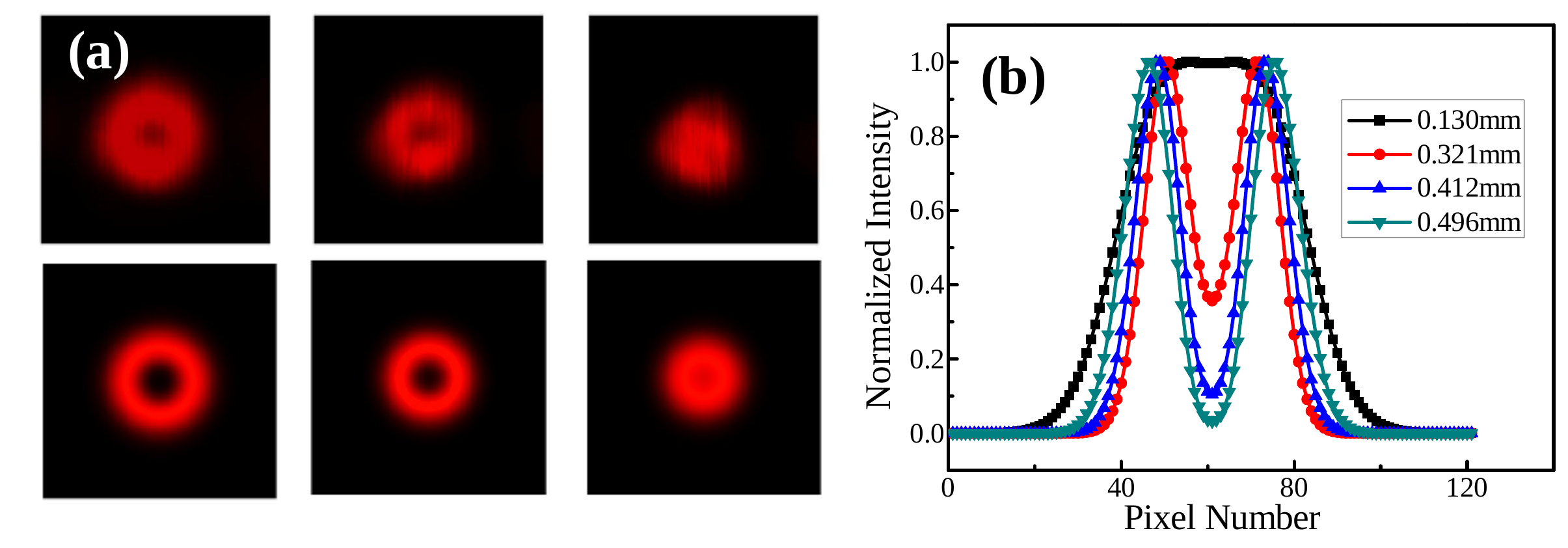}
       \caption{(Colour online) (a) The experimental (top) and theoretical (bottom) far field intensity distributions of the scattered LG beam of azimuthal index ($\rm m = 1$) through a RGG plate for different $\sigma$ values in \cite{korotkovaoe}, 0.496 \rm{mm} (left), 0.412 \rm{mm} (middle) and 0.140 \rm{mm} (right). (b) The line profiles of theoretical far field intensity distributions of scattered first order vortex for different incident beam widths.}
        \label{fig:line1} \vspace{0.1cm}
        \end{center}
      \end{figure}

  \begin{center}
  \begin{align} 
  A = 1 - \frac{z_2}{f}, \hspace{0.5cm}  B = z_1(1 - \frac{z_2}{f}), \nonumber \\ C = \frac{-1}{f}, \hspace{0.5cm}  D = \frac{-z_1}{f}.
  \end{align}
   \end{center}
   The results are in good agreement with the experimental results shown in Fig. \ref{fig:lgsm}. To quantify the disappearance of the dark core with the decreasing width of incident LG beam in our theoretical plots, we have given the intensity distributions for first order scattered vortex for different incident beam widths in Fig.\ref{fig:line1} (a) and shown that the dark core completely  disappears if the beam width is less than $0.140 \rm{mm}$. The line profiles of far field intensity distributions for first order scattered vortex shown in Fig. \ref{fig:line1} (b). From these line profiles also, it is clear that the dark core disappears if incident beam width is less than 0.130 \rm{mm}. One can obtain the similar results for the scattered BG beams also \cite{bcb}, as BG beams can be represented by shifted Hermite-Gaussian beams \cite{BG}. 

\begin{figure}[htb]
 \begin{center}
 \includegraphics[width=8.4cm]{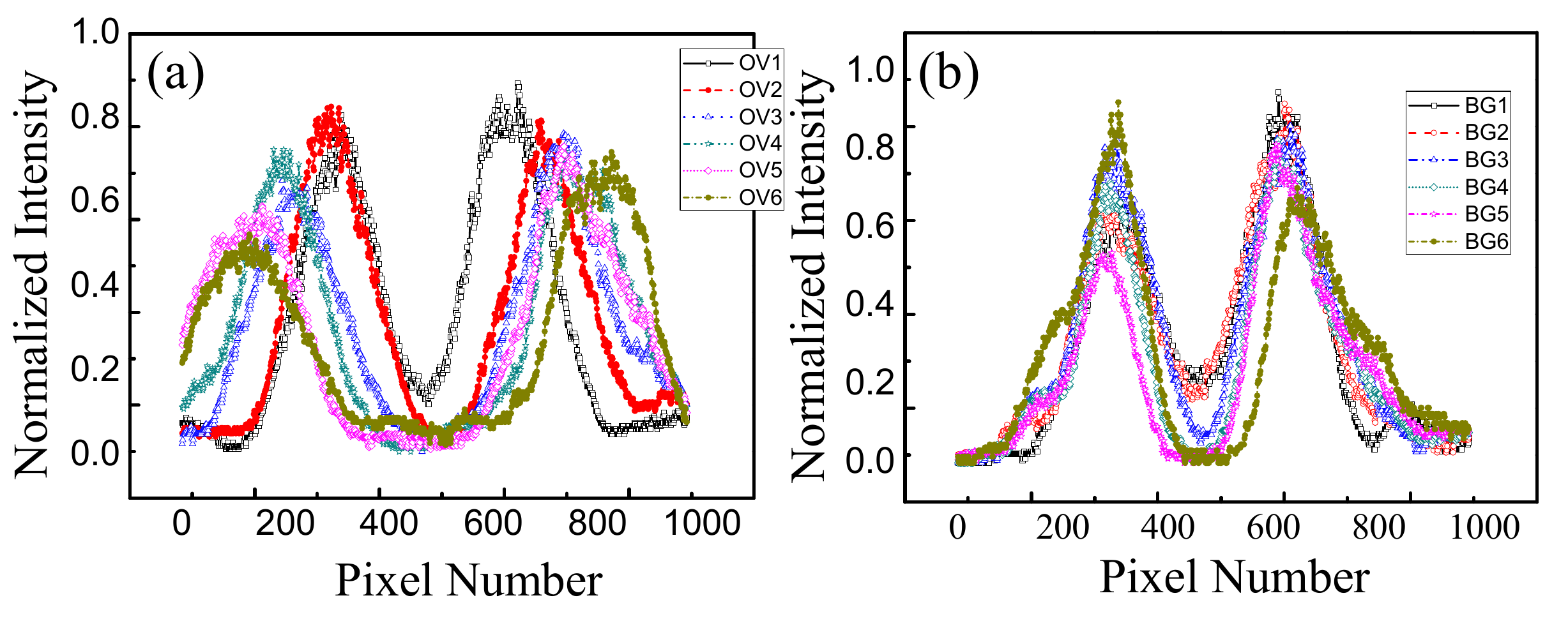}
  \caption{(Colour online) The line profiles (intensity distribution) along the centers of far field intensity distributions of scattered LG (a) and BG (b) beams for m = 1-6}
   \label{fig:line} \vspace{0.1cm}
   \end{center}
 \end{figure}

 We have also studied the effect of azimuthal index on the size of dark core of ring shaped beams at a given temporal and spatial coherence. We have plotted the line profiles through the centers of ring shaped beams formed by scattered LG and BG beams with different azimuthal indices (m = 1-6) for the incident beam size of 1.1 mm and the RGG speed of 34.3 cm/sec; shown in Fig. \ref{fig:line}. The dark cores of the ring shaped beams are quite prominent and as the azimuthal index increases it becomes broader.
 
\begin{figure}[h]
\begin{center}
\includegraphics[width=7.4cm]{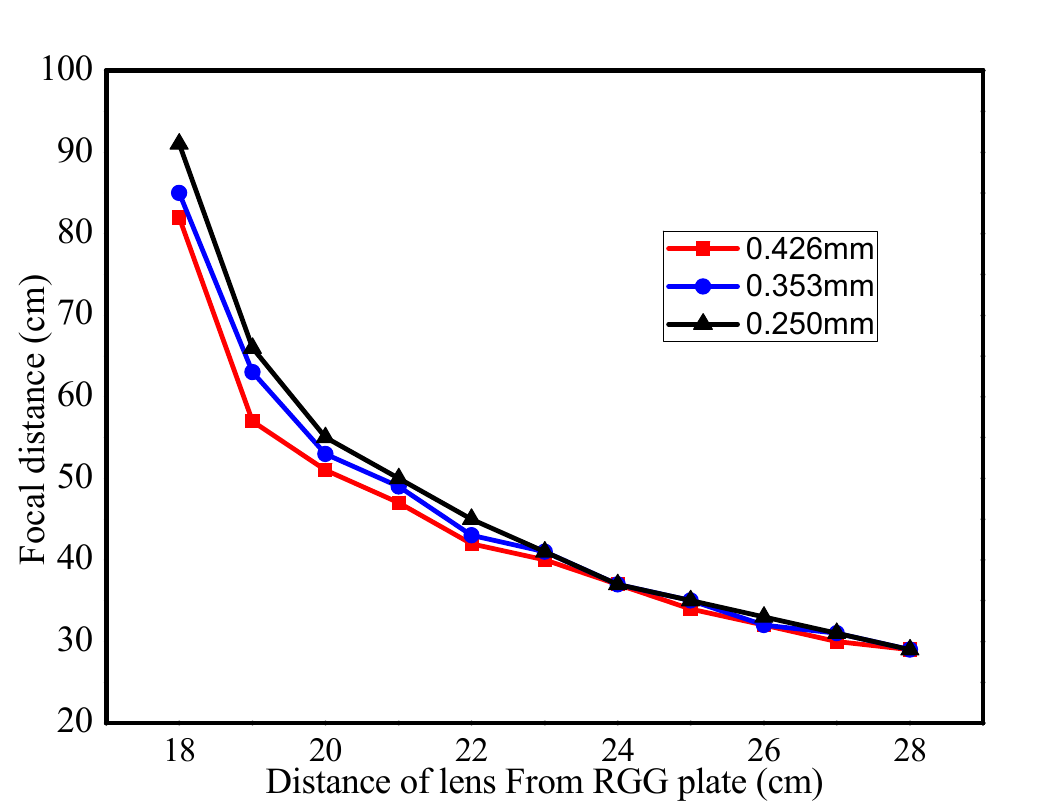}
 \caption{(Colour online) Plot showing the shifting of focal plane with the position of lens from the RGG plate for different beam widths.}
 \label{fig:focal}
 \end{center}
\end{figure}

In the course of our study on formation of the ring shaped beams, we observed a shift in the focal plane of the lens when it was moved away or towards the RGG plate. A plot between the position of focal plane (focal shift) and the distance of lens from the RGG plate has been shown in Fig. \ref{fig:focal}. It is clear, when the lens moves towards the RGG plate, the focal plane shifts away from the lens and vice versa. Similar results have been obtained by changing the aperture size placed in front of the lens for the GSM beams \cite{olga1}. In both the cases, the focal plane moves towards the lens when we collect less amount of partially coherent light or effectively decrease the aperture size. The effect of spatial coherence on focal shift has also been studied with the change of incident beam width. We have observed that at a given distance of lens from the RGG plate, the focal point shifts towards the lens with decrease in the beam width. Thus, we could say that with decrease in the spatial coherence of partially coherent light, the focal point of a lens shifts towards the lens.  

We have experimentally generated the ring-shaped beams by collecting the scattered light of LG and BG beams. We have also studied the dependence of ring-shaped beams on the spatial and the temporal coherence of light. The focal shift for partially coherent light beams has also been investigated. The generated ring-shaped beams may be of importance in optical trapping experiments. Also the controlled focal shift obtained may be useful in changing the trapping planes. In inverted optical tweezer set up, the focal plane shifts due to the refractive index mismatch between the immersion oil and cover slip \cite{trap}; such shifts may be compensated with controlled focal shifts in the path of trapping beams. The use of these beams for optical trapping experiments were preferable at higher speeds of the RGG plate as the beams get more and more smooth. 

\newpage

\pagebreak
\section*{Informational Fourth Page}

\end{document}